\documentclass[aps,pre,twocolumn,groupedaddress,showpacs,floatfix]{revtex4-1}

\usepackage[T1]{fontenc}
\usepackage[latin9]{inputenc}
\usepackage{booktabs}
\usepackage{ltxtable}
\usepackage{textcomp}
\usepackage{tabularx}
\usepackage{graphicx, color}
\usepackage{subfigure}
\usepackage{amsmath}
\usepackage{amsfonts}
\usepackage{amssymb}
\usepackage{amsthm}
\usepackage{empheq}
\usepackage{fixmath}
\usepackage{tensind}
\usepackage{wasysym}  %\th  needs encoding T1  \th = thorn operator   
\usepackage{afterpage}
\usepackage{dsfont}   % unit matrix symbol (\mathds{1}), or real numbers \mathds{R} ... 
\usepackage{hyperref}
\usepackage{makeidx}
\usepackage{textcomp}

\usepackage{color}
\usepackage{tikz}
\usepackage{tkz-berge}
\usetikzlibrary{matrix}
\usetikzlibrary{positioning,fit}
\usetikzlibrary{shapes,arrows}
\usetikzlibrary{positioning}
\usetikzlibrary{shapes}
\usetikzlibrary{snakes}

\usetikzlibrary{arrows,decorations.pathmorphing,backgrounds,positioning,fit,petri}

\pgfdeclarelayer{background}
\whenindex{w}{\mathbf{a}}{}
\whenindex{x}{\mathbf{b}}{}
\whenindex{y}{\mathbf{c}}{}
\whenindex{z}{\mathbf{d}}{}
\whenindex{W}{\mathbf{A}}{}
\whenindex{X}{\mathbf{B}}{}
\whenindex{Y}{\mathbf{C}}{}
\whenindex{Z}{\mathbf{D}}{}
\whenindex{'}{'}{}

\def\bsm{\left( \!\begin{smallmatrix}}
\def\esm{\end{smallmatrix} \!\right)}

\tensordelimiter{?}

\numberwithin{equation}{section}

\newcommand{\Z}{\mathbb{Z}}

\renewcommand{\i}{\mathrm i}

\newcommand{\be}{\begin{equation}}
\newcommand{\ee}{\end{equation}}
\newcommand{\bel}[1]{\begin{equation}\label{#1}}
\newcommand{\bea}{\begin{eqnarray}}
\newcommand{\eea}{\end{eqnarray}}
\newcommand{\bef}{\begin{figure}}
\newcommand{\enf}{\end{figure}}
\newcommand{\ba}{\begin{array}}
\newcommand{\ball}{\begin{array}{ll}}
\newcommand{\bacl}{\begin{array}{cl}}
\newcommand{\bacll}{\begin{array}{cll}}
\newcommand{\bal}{\begin{array}{l}}
\newcommand{\bac}{\begin{array}{c}}
\newcommand{\ea}{\end{array}}

\newcommand{\ie}{i.e.\, }

\definecolor{red}{rgb}{0.7,0.5,0.0}

\begin{document}

\bibliographystyle{apsrev}

\title{Collective Relaxation Dynamics of Small-World Networks} 

\author{Carsten Grabow${}^{a,b}$}%
\author{Stefan Grosskinsky${}^{c}$}%
\author{J\"urgen Kurths${}^{a,d,e}$}%
\author{Marc Timme${}^{b,f,g}$}%

\affiliation{%
${}^a$Research Domain on Transdisciplinary Concepts and Methods, Potsdam Institute for Climate Impact Research, P.O. Box 60 12 03, 14412 Potsdam, Germany\\
${}^b$Network Dynamics, Max Planck Institute for Dynamics and Self-Organization (MPIDS), 37077 G\"{o}ttingen, Germany\\
${}^c$Mathematics Institute and Centre for Complexity Science, University of Warwick, Coventry CV4 7AL, UK\\
${}^d$Department of Physics, Humboldt University of Berlin, Newtonstr. 15, 12489 Berlin, Germany \\
${}^e$Institute for Complex Systems and Mathematical Biology, University of Aberdeen, Aberdeen AB24 3UE, UK\\
${}^f$Institute for Nonlinear Dynamics, Faculty for Physics, Georg August University G\"{o}ttingen, 37077 G\"{o}ttingen, Germany\\
${}^g$Bernstein Center for Computational Neuroscience G\"{o}ttingen, 37077
G\"{o}ttingen, Germany}%

\date{\today}

\begin{abstract}
Complex networks exhibit a wide range of collective dynamic phenomena,
including synchronization, diffusion, relaxation, and coordination
processes. Their asymptotic dynamics is generically characterized by the local
Jacobian, graph Laplacian or a similar linear operator. The structure of
networks with regular, small-world and random connectivities are reasonably
well understood, but their collective dynamical properties remain largely
unknown. Here we present a two-stage mean-field theory to derive analytic
expressions for network spectra. A single formula covers the spectrum from regular via small-world to strongly
randomized topologies in Watts-Strogatz networks, explaining the simultaneous dependencies on
network size $N$, average degree $k$ and topological randomness $q$. 
We present simplified analytic predictions for the second largest and smallest eigenvalue, and numerical checks confirm our
theoretical predictions for zero, small and moderate topological randomness
$q$, including the entire small-world regime. For large $q$ of the order of
one, we apply standard random matrix theory thereby overarching the full
range from regular to randomized network topologies. These results may contribute to our analytic and mechanistic understanding of collective relaxation phenomena of
network dynamical systems. 
\end{abstract}


\maketitle

\section{Introduction}\label{relax}

The structural features of complex networks underlie their collective dynamics
such as synchronization, diffusion, relaxation, and coordination processes
\cite{SynchBook,Strogatz01}. Such processes occur in various fields, ranging
from opinion formation in social networks \cite{Pluchino05} and consensus
dynamics of agents \cite{Olfati2005} to synchronization in biological circuits
\cite{Buchanan:2010uf,   Manrubia:2004uv} and oscillations in gene
regulatory networks and neural circuits \cite{McMillen:2002fw,Gardner03,Hoppensteadt:1607706}. The asymptotic collective dynamics of all such
processes is characterized by the local Jacobian, the graph Laplacian or
similar linear operators. In general, such linear operators directly connect the structure of an underlying network to its dynamics and thus its function (see
e.g. \cite{Bollobas:0m7inuPk,Biyikoglu:2007ui}). Therefore, a broad area of
research is related to the study of properties of such operators, in
particular to the study of their spectra \cite{Motter:2005p76,Chung:2005ub,Motter:2005vk,AGAEV:2005io,Donetti:2006ck,Samukhin:2008hb,Jost08,Mcgraw:2008cu}. 

Small-world models based on rewiring have received widespread attention both
theoretically and in applications, as demonstrated, for instance, by the huge
number of references pointing to the original theoretical work
\cite{Watts:1998vz}. But for most of their features analytical predictions are
not known to date (\cite{weigt2000}; a mean field solution of its average path
length constitutes a notable exception \cite{Newman:2000vd}). In particular,
the spectrum of small-world Laplacians has been studied for several specific
cases and numerically
\cite{Monasson:1999vo,Jost01,Barahona:2002bm,Mori04,Kuhn11} yet a general derivation of reliable analytic predictions was missing. 

Here we present a mean field theory \cite{Grabow:2012dg} towards closing this gap. The article is organized as follows. In Section \ref{relax} we first review
the relations between relaxation dynamics and the spectrum of the graph
Laplacian. In Section \ref{mftrewire} we present rewiring `on average', a
new mean field rewiring recently proposed in a brief report
\cite{Grabow:2012dg}. Based on this rewiring, we derive a single formula that
approximates well the entire spectrum from regular to strongly randomized topologies. 
We then investigate the ordering of the mean field
eigenspectrum in Section \ref{ordering}. In Section \ref{extreme} we quantify the accuracy of our predictions via systematic numerical checks for the extreme eigenvalues. For the topological randomness $q$ of the order of unity, standard random matrix theory is applied in Section \ref{RMT}. We close in Section \ref{sum2} with a summary and a discussion of further work.

\section{Network Relaxation Dynamics}\label{relax}

The relaxation dynamics towards equilibrium and related collective phenomena close to
invariant sets or stationary distributions emerge across a wide variety of systems ubiquitous in natural and artificial systems \cite{Arenas:2008p1192, SynchBook}.

\subsection{Generic linear relaxation}
Mathematically, the relaxation of network dynamics to a fixed point,
a periodic orbit or a similar stationary regime is generically characterized by
equations of the form

\begin{equation}
\frac{d x_{i}}{dt}= \sum_{j=1}^{N}J_{ij}(x_{j}-x_{i}) \hspace{0.5cm}\,\mbox{for}\, i\in\{1,...,N\}\ ,
\label{eq:generic}
\end{equation}
where $x_i(t)$ quantifies the deviation at time $t$ from an invariant state, $N\in\mathbb{N}$ is the size of the network
and $J_{ij}\in\mathbb{R}$ quantifies the influence of unit $j$ onto unit $i$. In general $x_i(t)\in\mathbb{R}^d$, we here
take $d=1$ for simplicity of presentation.

The equivalent mathematical description 
\begin{equation}
\frac{d x_{i}}{dt}= \sum_{j=1}^{N}\Lambda_{ij} x_{j} 
\label{eq:genericLaplacian}
\end{equation}
for $ i\in\{1,...,N\}$ with the Laplacian
\begin{equation}
\Lambda_{ij}=
\left\{ 
\begin{array}{ll}
J_{ij} & \mbox{for}\  i\neq j  \\[2mm]
-\sum_{j=1}^N J_{ij}& \mbox{for}\  i =j 
\end{array}
\right .  
\label{eq:LaplacianDef}
\end{equation}
follows directly from the original dynamics (\ref{eq:generic}).

The eigenvalues $\lambda_n\in\mathbb{C}$ and corresponding eigenvectors $v_n$ of such a
Laplacian, satisfying 
\begin{equation}
\Lambda v_n = \lambda_n v_n
\label{eq:LaplacianEVs}
\end{equation}
for $n \in \{1,\ldots,N \}$, fully characterize the asymptotic (linear or
linearized) dynamics. For instance, for stable dynamics, where all $x_i(t) \rightarrow 0$
for $t \rightarrow \infty$, the largest
nonzero (principal) eigenvalue $\lambda_*$ dominates the long term dynamics: if we have
$x_j(0)=\sum_{n=1}^N a_n v_n$, the vector
$x(t)=(x_1(t),\ldots,x_N(t))^\mathsf{T}$ evolves as  

\begin{eqnarray} \label{eq:xEvolution}
x(t) & = & \exp(\Lambda t)x(0) \\ \nonumber
& = & \exp(\Lambda t) \sum_{n} a_n  v_n \\ \nonumber
& = &  \sum_{n=1}^{N} a_n \exp (\lambda_n t) v_n .
\end{eqnarray}
Due to stability we have $a_n=0$ whenever $\lambda_n=0$, and for long times this is dominated by 
\begin{equation}
x(t) \sim a_* \exp (\lambda_* t) v_* .
\label{eq:xAsymptotic}
\end{equation}

\noindent where $\lambda_*$ is the principal eigenvalue. 

Note that (\ref{eq:xEvolution}) also reveals how exactly all
eigenvalues contribute to relaxation (and how much relative to each
other). Additional individual eigenvalues of interest are given by the one
with smallest real part $\lambda_-$, because it bounds the real parts of
the spectrum below and thereby determines the support of the spectrum, and
also because it is involved in determining synchronizability conditions in
coupled chaotic systems via the ratio $\lambda_*/\lambda_-$, see \cite{Pecora:1998up}.
 
\subsection{Different example systems}
 
We  briefly consider two very different paradigmatic example classes of
systems and comment on a few others whose
relaxation properties are characterized by equations of the same type as (\ref{eq:genericLaplacian}).

\emph{Stochastic processes.} Firstly, consider random walks on a graph, or equivalently, Markov
chains defined by a weighted non-negative graph whose nodes represent the
$N$ states \cite{Kriener:2012ib}.
For such processes, the dynamics of the probability
$p_i(t)$ to reside state $i \in \{1,\ldots,N\}$ at time $t$ is given by
\begin{equation}
\frac{d p_{i}}{dt}= \sum_{j=1}^{N}\left[T_{ij}p_{j}-T_{ji}p_{i}\right] \hspace{0.5cm}\,\mbox{for}\, i\in\{1,...,N\}\ ,
\label{eq:genericMarkov}
\end{equation}
where $T_{ij}$ defines the transition rate (probability per unit time) of the
system switching state from $j$ to $i$ given it resides in $i$.
We assume the process to be ergodic. Identifying $p^{*}$ to be the unique stationary distribution $\Lambda_{ij}=T_{ij}$ for $i\neq j$ and $\Lambda_{ii}=-\sum_jT_{ji}$ and setting $x_i(t) \equiv p_i(t) - p^{*}_{i} $ exactly maps this processes
onto the generic form (\ref{eq:genericLaplacian}) with $x_i(t) \rightarrow 0$
as $t \rightarrow \infty$ for all $i \in \{1,\ldots,N\}$.

\emph{Coupled deterministic oscillators.} Secondly, consider the relaxation dynamics of weakly coupled limit cycle
oscillators, generically modeled as phase-coupled oscillators
\begin{equation}
\frac{d\theta_{i}}{dt}=\omega_{i}+ \sum_{j}h_{ij}(\theta_{j}-\theta_{i}) \hspace{0.5cm}\,\mbox{for}\, i\in\{1,...,N\}\ ,
\label{eq:phaseoscillators}
\end{equation}
where $\theta_i(t)$ is the phase of unit $i$ at time $t$, $\omega_i$ is the
local intrinsic frequency of oscillator $i$ and $h_{ij}(.)$
defines a smooth coupling function from unit $j$ to $i$ \cite{Hoppensteadt:1607706,Kuramoto:vq,Acebron:2005p293}. 
Phase-locking, where $\theta_{j}(t)-\theta_{i}(t)=\Delta_{ji}$ is constant in time, constitutes
a generic collective state of such systems, see e.g.~\cite{Kaka:2005gj,Timme:2007fs,Grabow10}.
A paradigmatic model is given by networks of Kuramoto oscillators coupled by
simple sinusoidal functions, i.e.~ $h_{ij}(\theta)=\sin(\theta)$ \cite{Kuramoto:vq,Acebron:2005p293,Witthaut:2014cd}. 
	
In the most general setting, a matrix $J$ is defined by elements
$J_{ij} = \partial h_{ij} (\theta )/\partial\theta |_{\theta =\Delta_{ji}}$ that encode
the network structure close to the phase-locked state. Under certain conditions on the
$\omega_i$ and the $h_{ij}$, the system's dynamics exhibits a short transient
dominated by nonlinear effects and thereafter exponentially relaxes to the
phase-locked state. Linearizing close to such a state yields
phase perturbations defined as
\begin{equation}
\delta_{i}(t):=\theta_{i}(t)-\theta(t)
\end{equation}
evolve according to
\begin{equation}
\frac{d\delta_{i}}{dt}= \sum_{j}\Lambda_{ij}\delta_{j}(t)\hspace{0.5cm}\,\mbox{for}\, i\in\{1,...,N\} \,\label{linearmodel}
\end{equation}
with the graph Laplacian given by (\ref{eq:LaplacianDef}).

In a simple setting, we have  $J_{ij}=1$  for an
existing edge and $J_{ij}=0$ for no edge such that the local linear operator in (\ref{linearmodel}) coincides with the graph Laplacian defined by its elements
\begin{equation}
\Lambda_{ij}=J_{ij}(1-\delta_{ij}) - k_{i} \delta_{ij}  \label{eq:laplacianinmftsec}
\end{equation}
\noindent for $i,j \in \{1,\dots,N\} $, where now $J_{ij}$ are the elements of
the adjacency matrix, $k_{i}$ is the degree of node $i$ (replaced by the
in-degree for directed networks) and $\delta_{ij}$ is the Kronecker-delta. 
The asymptotic relaxation dynamics on such networks is thus characterized by
this graph Laplacian $\Lambda$. Similarly, any dynamics near genuine fixed
points, for instance in gene regulatory networks \cite{collins03,Timme:2014hy} 
is equally characterized by linearized dynamics stemming from local Jacobians.

\emph{Power grids, social networks, neural circuits,...} Several other systems exhibit qualitatively
the same dynamical relaxation. In fact, power grids are often characterized by
second order oscillatory systems \cite{12powergrid} 
that principally relax to periodic phase-locked solutions
(stationary operating states of the grid) very similarly to the phase oscillator
systems discussed above \cite{Manik:2014gea}. 
In models of several social phenomena, e.g. of opinion formation of agents, the dynamics of consensus
formation is equally akin to such locking dynamics, where the
locked state would now be a homogeneous, fully synchronous one \cite{OlfatiSaber:2007cma}. 
Last but not least, perturbations of the spatio-temporal
collective dynamics of pulse-coupled systems such as neural circuits \cite{Timme04,Timme06} also relax according to (\ref{eq:genericLaplacian}).

\section{Mean field rewiring and spectrum}\label{mftrewire}

Diving into explaining the small-world model, we analyze
and derive its approximate Laplacian based on a two stage mean field rewiring. We follow \cite{Grabow:2012dg} and where appropriate take parts of the description presented there. 

\begin{figure}[t]
\begin{center}
\includegraphics[width=85mm]{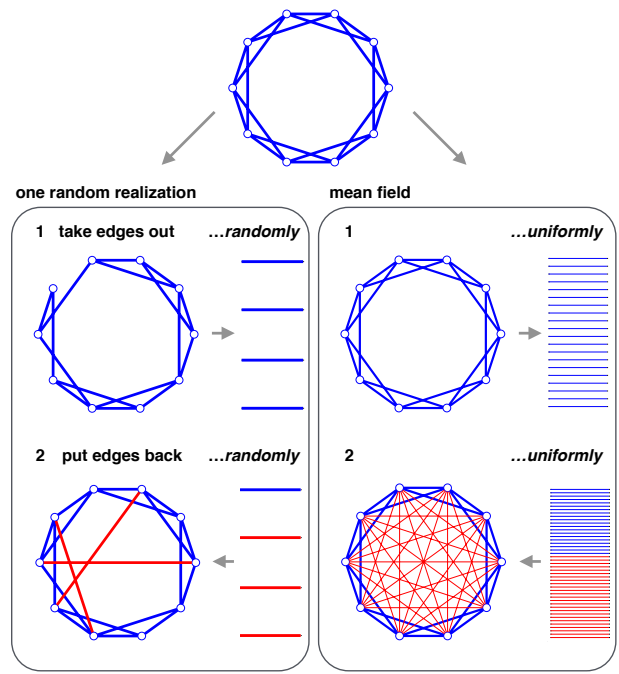}
\caption[Two-stage mean field rewiring.]{\label{fig:networks}
  \textbf{Rewiring -- stochastic and mean field.} (Cartoon for $N=10$ and $k=4$)
  Instead of taking out (step \textbf{1}) and putting back (step \textbf{2})
  edges randomly (left column) the corresponding weight is subtracted (step
  \textbf{1}) uniformly and added (step \textbf{2}) in two fractions (right
  column).
}
\end{center}
\end{figure}

We consider an initial ring graph of $N$ nodes. Each node receives $k$ (being even) links from its $k/2$ nearest neighbors
on both sides. Then we introduce randomness in the network topology by rewiring. 

To define Watts-Strogatz randomized networks,
single instances of an ensemble of stochastically rewired networks are
generated (Fig.~\ref{fig:networks}, left panel). 
Following \cite{Watts:1998vz} for undirected networks, we first cut each edge with probability $q$. Afterwards the cut edges are rewired to nodes chosen uniformly at random from the whole network. Similarly, for directed \cite{Fagiolo:2007th} networks, we first cut all out-going edges with probability $q$ and rewire their tips afterwards. In both cases we avoid double edges and self-loops.

To analytically determine the Laplacian mean field spectrum in dependence of the network size $N$, the average degree $k$ and the topological randomness $q$, we introduce a two-stage mean field rewiring that effectively generates, at given $q$ the average network from the ensemble of all stochastically rewired networks. This is depicted in Fig.~\ref{fig:networks} (right panel) in comparison to both other rewiring procedures for undirected and directed networks. Firstly, we define a circulant mean field Laplacian
\begin{equation}
\tilde{\Lambda}^{\textrm{mf}}=\begin{pmatrix}  c_{0}	  & c_{1} & c_{2}    &               & \cdots & c_{N-1} \\
			      c_{N-1} & c_{0} & c_{1}   & c_{2}     &  		& \vdots  \\ 
			      		  & c_{N-1} & c_{0} & c_{1}     & \ddots & 		 \\
			      \vdots   & 	          & \ddots & \ddots   & \ddots & c_{2} \\
			      		  &  &  &         &  & c_{1} \\
				c_{1} &  \cdots &  &         & c_{N-1} & c_{0}   \end{pmatrix} \ .
				\label{eq:mftlaplacian}
\end{equation}
Its matrix elements for the initial configuration (Fig.~\ref{fig:networks}, q=0, top) are given by
\begin{equation}
c_{i}=\left\{ \begin{array}{ll}
-k & \mbox{if}\  i = 0 \\
1 & \mbox{if}\  i \in \{1,\hdots,\frac{k}{2},N-\frac{k}{2},\hdots,N-1\}=S_{1} \\
0 & \mbox{if}\ i \in \{\frac{k}{2}+1,\hdots, N-\frac{k}{2}-1\}=S_{2} \ , \end{array}\right .  \label{eq:ringmatrixelements}
\end{equation}
where $S_{1}$ represents the set of edges being present in the ring and $S_{2}$ those absent ones outside that ring. 

Instead of rewiring single edges randomly (Fig.~\ref{fig:networks}, left panel), we now distribute the corresponding weight of rewired edges uniformly among the whole network (Fig.~\ref{fig:networks}, right panel). Thus, for a given rewiring probability $q$ we generate a mean field version of the randomized network ensemble in the following two steps:

Firstly, we subtract the average total weight $q k N/2$ of all edges to be rewired ($S_{1}$), \ie $c_{i}=1-q$ if $i \in S_{1}$.

\begin{figure} [t] 
\begin{center}
\includegraphics[width=85mm]{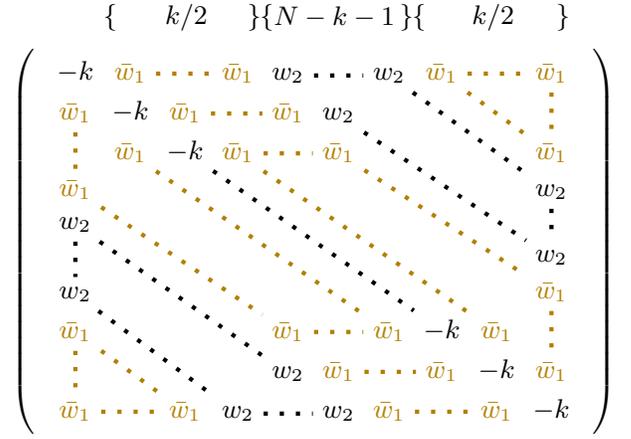}
\caption[The banded structure of the mean field graph Laplacian $\tilde{\Lambda}^{\textrm{mf}}$ given in Eqs.~(\ref{eq:mftlaplacian}) and (\ref{eq:eigenmatrixelements}).]{\textbf{The banded structure of the mean field graph Laplacian $\tilde{\Lambda}^{\textrm{mf}}$ given in Eqs.~(\ref{eq:mftlaplacian}) and (\ref{eq:eigenmatrixelements}).} It has the weights $\bar{w}_{1}=1-q+w_{1}$ for $c_{i} | i \in S_{1}$ and $w_{2}$ for $c_{i} | i \in S_{2}$ (see Eq.~(\ref{eq:ringmatrixelements}) for the definition of $S_{i}$). For $q=0$ and hence $\bar{w}_{1}=1$ and $w_{2}=0$ we recover the exact ring Laplacian.} \label{fig:mftlaplacian}
\end{center}
\end{figure}

Secondly, the rewired weight is distributed uniformly among the total 'available' weight in the whole network given by 
\bel{eq:weighttotal}
f=\frac{N(N-1)-(1-q) k N}{2} .
\ee
With the weights 
\be
f_{1}=\frac{q k N}{2}
\ee being available in $S_{1}$ and 
\be
f_{2}=\frac{N(N-1)- k N}{2}
\ee in $S_{2}$, we assign the fraction $f_{1}/f$ to elements representing edges in $S_{1}$ and $f_{2}/f$ to those representing $S_{2}$. 
Therefore, an individual edge in $S_{1}$ gets the additional weight 
\bel{eq:weight1}
w_{1}= \frac{f_{1}}{f} \frac{\frac{qkN}{2}}{\frac{kN}{2}}=\frac{q^{2}k}{N-1-(1-q)k} \ ,
\ee
and an edge in $S_{2}$ gets the new weight
\bel{eq:weight2}
w_{2}= \frac{f_{2}}{f} \frac{\frac{qkN}{2}}{\frac{N(N-1)-kN}{2}}=\frac{qk}{N-1-(1-q)k} \ .
\ee
Thus, as depicted in Fig.~\ref{fig:mftlaplacian}, in our mean field theory the elements of the Laplacian $\tilde{\Lambda}^{\textrm{mf}}$ (\ref{eq:mftlaplacian}) of a network on $N$ nodes with degree $k$ after rewiring with probability $q$ are given by
\begin{equation}
c_{i}=\left\{ \begin{array}{ll}
-k & \mbox{if}\  i=0 \\
1-q+ w_{1} & \mbox{if}\   i \in S_{1} \\
w_{2} & \mbox{if}\ i \in S_{2} \ . \end{array}\right.\label{eq:eigenmatrixelements}
\end{equation}

The mean field Laplacian defined by (\ref{eq:mftlaplacian}) and (\ref{eq:eigenmatrixelements}) by construction is a circulant matrix with eigenvalues \cite{circbook,matrices85,gray01}
\bel{eq:mflaplacian}
\tilde{\lambda}^{\textrm{mf}}_{l}=\sum^{N-1}_{j=0}c_{j} \exp{\left(\frac{-2\pi \i (l-1) j}{N}\right)} \ .
\ee

Observing the structure in Fig.~\ref{fig:mftlaplacian} we immediately obtain the trivial eigenvalue for $l = 1$:
\begin{equation}
\tilde{\lambda}^{\textrm{mf}}_{1} =  \sum^{N-1}_{j=0} c_{j} = -k + k (1 - q + w_{1}) + (N-k-1) w_{2} = 0 \ , 
\end{equation}
which is common to all networks (for all $q$, $N$, and any $k\leq N-1$) and reflects the invariance of Laplacian dynamics against uniform shifts, as seen from the associated eigenvector $\tilde{v}_1=(1,\ldots,1)^{\textsf{T}}$. 

To obtain the remaining eigenvalues for $l \in \{2,\dots,N\}$ we first define
\be
x_{l}:=\exp{\left(\frac{-2\pi \i (l-1)}{N}\right)} \ ,
\ee
\be
c^\prime:=\frac{1- q}{k} + q c^{\prime\prime}
\ee
and
\be
c^{\prime\prime}:=\frac{q}{N-1-(1-q)k} \ .
\ee
This leads to \begin{align}
\tilde{\lambda}^{\textrm{mf}}_{l}=  & -k +  k c^\prime \sum^{\frac{k}{2}}_{j=1} x_{l}^{j} \nonumber \\
& + k c^{\prime\prime} \sum^{N-1-\frac{k}{2}}_{j=\frac{k}{2}+1} x_{l}^{j} 
+ k c^\prime \sum^{N-1}_{j=N-\frac{k}{2}} x_{l}^{j} \\
= & -k + k c^\prime \sum^{\frac{k}{2}}_{j=1} x_{l}^{j} 
+ k c^\prime \sum^{\frac{k}{2}}_{j=1} x_{l}^{N-j} \nonumber \\
&+ k c^{\prime\prime} \left(\sum^{\frac{N-k}{2}-1}_{j=1}x_{l}^{\frac{N}{2}+j}+\sum^{\frac{N-k}{2}-1}_{j=1} x_{l}^{\frac{N}{2}-j}+x_{l}^{\frac{N}{2}}\right) \ , \end{align}
where we have exploited the additional transposition symmetry $\tilde{\Lambda}^{\textrm{mf}}=\left(\tilde{\Lambda}^{\textrm{mf}}\right)^{\textsf{T}}$ which implies $c_{j}=c_{N-j}$.
Applying the Euler formula $\exp{\left(\i \alpha\right)}=\cos \left(\alpha\right) + \i \sin \left(\alpha\right) \ , $ the complex summands cancel and we get
\begin{align}
\tilde{\lambda}^{\textrm{mf}}_{l}= & -k +  2 k c^\prime \sum^{\frac{k}{2}}_{j=1} \cos\left(\frac{2\pi (l-1)j}{N}\right) \nonumber \\
& +  x_{l}^{\frac{N}{2}} k c^{\prime\prime} \left( 2 \sum^{\frac{N-k}{2}-1}_{j=1} \cos\left(\frac{2\pi (l-1)j}{N}\right) + 1 \right) \ .
\end{align}
\noindent Using the identity
\begin{align}
\sum_{j=0}^{n} \cos (j \alpha)=& \cos \left( \frac{n+1}{2} \alpha \right) \sin \left( \frac{n \alpha}{2} \right) \frac{1}{\sin(\frac{\alpha}{2})}+1\nonumber  \\
=&\frac{1}{2}\left( 1 + \frac{\sin\left((n+\frac{1}{2})\alpha\right)}{\sin(\frac{\alpha}{2})} \right) \ ,
\end{align}
we obtain
\begin{align}
\tilde{\lambda}^{\textrm{mf}}_{l} = &  -k + k c^\prime \left(\frac{\sin\left(\frac{(k+1)(l-1) \pi}{N}\right)}{\sin\left(\frac{(l-1)\pi}{N}\right)} -1\right) \nonumber \\
& + x_{l}^{\frac{N}{2}} k c^{\prime\prime} \frac{\sin\left(\frac{(N-k-1)(l-1)\pi}{N}\right)}{\sin\left(\frac{(l-1)\pi}{N}\right)} \ .\label{eq:l2with0} 
\end{align}

\begin{figure}[t]
\begin{center}
\includegraphics[width=85mm]{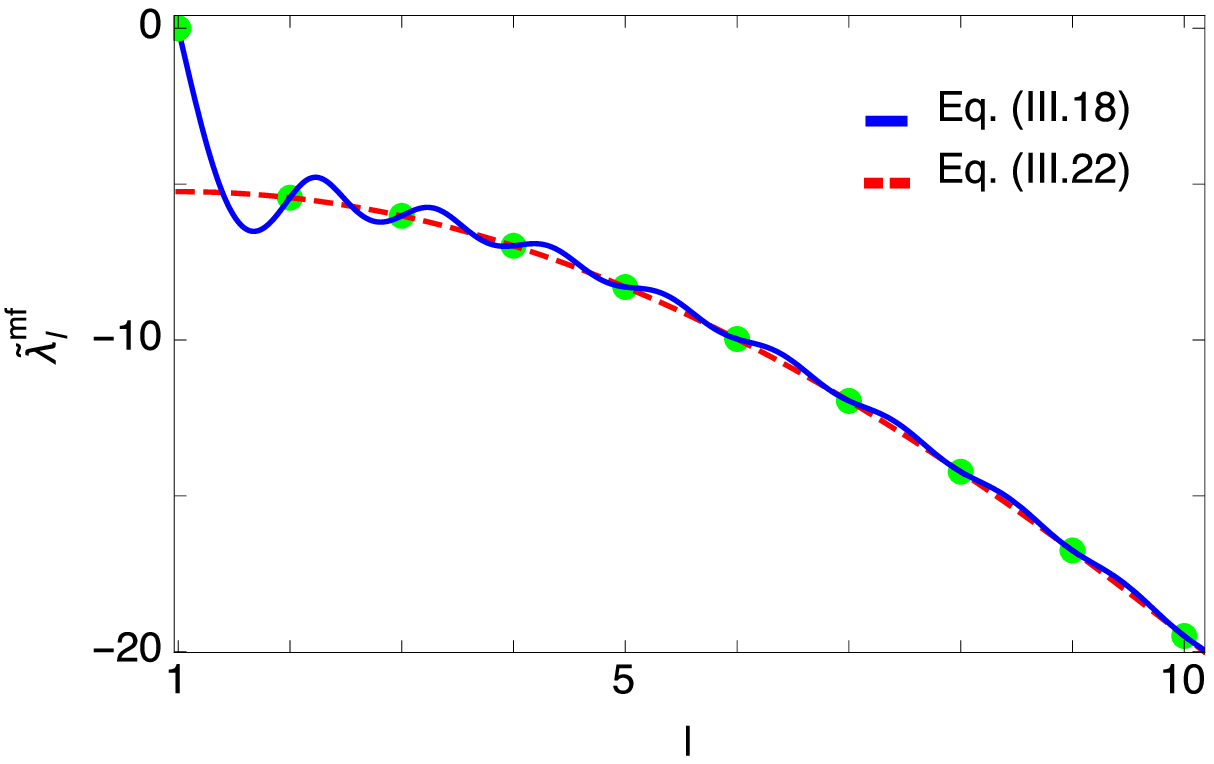}
\caption[Interpolating the eigenvalues.]{\label{fig:interplot} \textbf{Interpolating the eigenvalues.} Eqs.~(\ref{eq:l2with0}) (blue) and (\ref{eq:lambdafinal}) (red) both contain the eigenspectrum $\tilde{\lambda}^{\textrm{mf}}_{l}$ for $l \in \{2,\dots,N\}$ correctly (green circles). While $\tilde{\lambda}^{\textrm{mf}}_{l}$ in Eq.~(\ref{eq:l2with0}) includes $\tilde{\lambda}^{\textrm{mf}}_{1}$ as $\lim_{l \to 1} \tilde{\lambda}^{\textrm{mf}}_{l}=\tilde{\lambda}^{\textrm{mf}}_{1}=0$ as well, $\tilde{\lambda}^{\textrm{mf}}_{l}$ for $l\in\{2,\ldots ,N\}$ in Eq.~(\ref{eq:lambdafinal}) does not: To further simplify expressions, we have used identities (\ref{eq:id1}) and (\ref{eq:id2}) only valid for integer $l$, but apparently not for $l=0$.} 
\end{center}
\end{figure}

Taking advantage of additional identities -- only valid for $l \in \Z$ (Fig.~\ref{fig:interplot}) --
\bel{eq:id1}
x_{l}^{N/2}=(-1)^{l-1} \ ,
\ee
\bel{eq:id2}
(-1)^{l-1} \sin (\alpha) = \sin(\alpha + (l-1) \pi) 
\ee
and the symmetry $ \sin(-\alpha)=-\sin(\alpha)$,
the expression simplifies to
\begin{widetext}
\begin{align} 
\tilde{\lambda}^{\textrm{mf}}_{l} =  &-k + k c^\prime \left(\frac{\sin\left(\frac{(k+1)(l-1) \pi}{N}\right)}{\sin\left(\frac{(l-1)\pi}{N}\right)} -1\right)
+ (-1)^{l-1} k c^{\prime\prime} \frac{\sin\left(\frac{(-(k+1)+N)(l-1) \pi}{N}\right)}{\sin\left(\frac{(l-1)\pi}{N}\right)} \\ \nonumber
=  &-k + k c^\prime \left(\frac{\sin\left(\frac{(k+1)(l-1) \pi}{N}\right)}{\sin\left(\frac{(l-1)\pi}{N}\right)} -1\right)
+ k c^{\prime\prime} \frac{\sin\left(\frac{(-(k+1)(l-1)+2N(l-1)) \pi}{N}\right)}{\sin\left(\frac{(l-1)\pi}{N}\right)}  \\ \nonumber
=  &-k + k c^\prime \left(\frac{\sin\left(\frac{(k+1)(l-1) \pi}{N}\right)}{\sin\left(\frac{(l-1)\pi}{N}\right)} -1\right)
- k c^{\prime\prime} \frac{\sin\left(\frac{(k+1)(l-1) \pi}{N}\right)}{\sin\left(\frac{(l-1)\pi}{N}\right)} \ ,
\end{align}
\end{widetext}
which finally leads to
\bel{eq:lambdafinal}
\tilde{\lambda}^{\textrm{mf}}_{l} =   -k - k c^\prime + k \left(c^\prime -  c^{\prime\prime}\right) \frac{\sin\left(\frac{(k+1)(l-1) \pi}{N}\right)}{\sin\left(\frac{(l-1)\pi}{N}\right)} 
\ee
for $l \in \{2,\dots,N\}$.  

\section{The ordering of the mean field spectrum}\label{ordering}
The spectrum obeys the symmetry 
\bel{eq:specsym}
\tilde{\lambda}^{\textrm{mf}}_{l}=\tilde{\lambda}^{\textrm{mf}}_{N-l+2} \ ,
\ee
but is unordered otherwise, \ie the index $l$ does neither denote eigenvalues with decreasing real part nor
eigenvalues with decreasing absolute value.

\begin{figure}[t]
\begin{center}
\includegraphics[width=85mm]{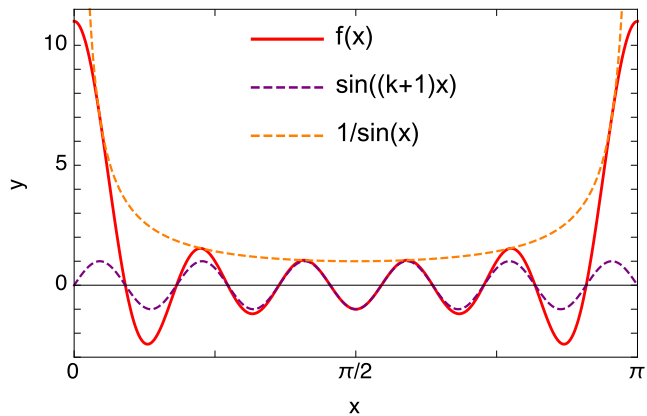}
\caption[$\tilde{\lambda}^{\textrm{mf}}_{2}$ always constitutes the second largest eigenvalue.]{\label{fig:fplot} \textbf{$\tilde{\lambda}^{\textrm{mf}}_{2}$ always constitutes the second largest eigenvalue.} Functions $f(x)$ (\ref{eq:f}), the oscillating function $\sin((k+1)x)$ and the envelope function $1/\sin(x)$ are plotted vs. $x= \frac{(l-1) \pi}{N} \in (0,\pi)$ for $k=10$. Obviously, a larger $k$ leads to more roots of $f(x)$, but otherwise the functions show the same characteristics for all $k\leq N-1$: $f(x)$ has a local maximum at $x=0$ and decreases strictly monotonically up to the following minimum. For larger $x$ the envelope function guarantees that all values up to $x = \pi/2$ are smaller than $f(x_{l=2} = \frac{\pi}{N})$.}
\end{center}
\end{figure}

As we argue below the expression $\tilde{\lambda}^{\textrm{mf}}_{2}$ (that equals $\tilde{\lambda}^{\textrm{mf}}_{N}$ due to (\ref{eq:specsym})) always constitutes the second largest (principal) eigenvalue $\tilde{\lambda}^{\textrm{mf}}_{*}$.
The only term depending on $l$ in eq. (\ref{eq:lambdafinal}) is the ratio
\be
\frac{\sin\left(\frac{(k+1) (l-1) \pi}{N}\right)}{\sin\left(\frac{(l-1)\pi}{N}\right)} \ .
\ee
We therefore study the function
\be
f(x)=\frac{\sin\left((k+1) x \right)}{\sin x } \label{eq:f} \ ,
\ee
with 
\be 
x= \frac{(l-1) \pi}{N} 
\ee
and $x \in (0,\pi/2)$. Due to the symmetry (\ref{eq:specsym}) the interval $(0,\pi/2)$ covers the entire spectrum (\ref{eq:lambdafinal}).	

The function $f(x)$ on $x \in (0,\pi/2)$ is the product of the
oscillating function $\sin((k+1)x)$ and a strictly
monotonically decreasing function $1/\sin(x)$. Therefore, it is a damped oscillation with period of $2 \pi /(k+1)$ and with the amplitude decreasing as $1/\sin(x)$ (Fig.~\ref{fig:fplot}). 

At $x=0$ we apply the Theorem of l'Hospital to calculate the following limits. There is a removable singularity 
\be
\lim_{x \to 0}f(x)=k+1 
\ee
with 
\be
\lim_{x \to 0}f'(x)=0 \ \mbox{and} \ \lim_{x \to 0}f''(x)=-\frac{1}{3}k(k^{2}+3k+2)<0 \ ,
\ee
i.e. a local maximum. 

\begin{figure}[t] 
\begin{center}
\includegraphics[width=85mm]{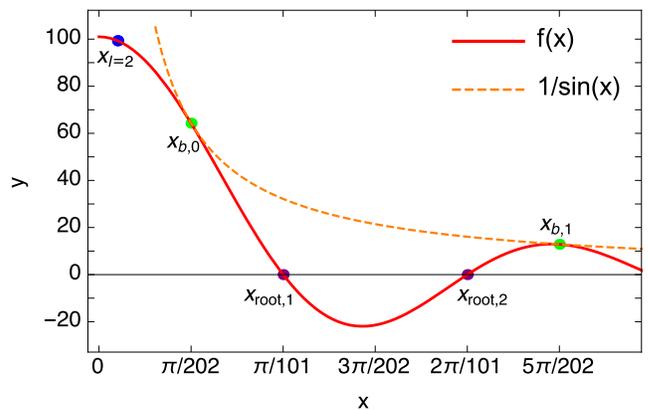}
\caption[Important points of the function $f(x)$]{\label{fig:fplot2} \textbf{Important points of the function $f(x)$.} The function $f(x)$ is plotted for $k=100$. The boundary points (green, eq. (\ref{eq:bpoints})) of the function $f(x)$ and the envelope function $1/\sin(x)$, roots of $f(x)$ (purple, eq. (\ref{eq:root})) and the x-value $x_{l=2}$ (blue, eq. (\ref{eq:xl2}), $N=1000$) corresponding to the eigenvalue $\tilde{\lambda}^{\textrm{mf}}_{2}$ are highlighted.
\label{fig:fplot2}}
\end{center}
\end{figure}

In order to show that the index $l=2$ is always associated with the second largest eigenvalue, we first determine its $x$-value. It is given by
\bel{eq:xl2}
x_{l=2}= \frac{\pi}{N} \ .
\ee

Since the roots of the function $f(x)$ are located at 
\bel{eq:root} 
x_{\text{root},r}=\frac{r \pi}{k+1} 
\ee 
for $r \in \Z$.
Thus, $x_{l=2}$ is always smaller than the first root $x_{\text{root},1}$ (\ref{eq:root}) of the function $f(x)$ (Fig.~\ref{fig:fplot2}).
 
The boundary points of function $f(x)$ and the envelope function $1/\sin(x)$ are given by
\bel{eq:bpoints}
x_{b,r}=\frac{4 \pi r + \pi}{2(k + 1)} \ .
\ee
for $ r \in \Z$. 

The function $f(x)$ is bounded from above by the envelope function $1/\sin(x)$ for all $x > x_{b,0}$ with 
\bel{eq:xb0}
x_{b,0}=\frac{\pi}{2(k + 1)}
\ee 
being the first boundary point of function $f(x)$ and its envelope function ($r=0$ in eq.~(\ref{eq:bpoints})) (Fig.~\ref{fig:fplot2}). 

The first derivative of $f(x)$ stays negative at least up to the first root ($r=1$ in eq.~(\ref{eq:root})) at 
\be
x_{\text{root},1} = \frac{\pi}{k+1} > x_{b,0}
\ee 
which is always larger than the first boundary point $x_{b,0}$ (\ref{eq:xb0}).

To summarize, the function $f(x)$ has a local maximum at $x=0$ and is then strictly monotonically decreasing up to $x_{b,0}$ (\ref{eq:xb0}). Then, for all $x>x_{b,0}$ the function $f(x)$ takes values smaller than or at most equal to the values of the envelope function $1/\sin(x)$, which is strictly monotonically decreasing in the considered domain (see Figures \ref{fig:fplot} and \ref{fig:fplot2}). Thus, if $x_{l=2}$ (\ref{eq:xl2}) is smaller than the first boundary point $x_{b,0}$ (\ref{eq:xb0}), the eigenvalue $\lambda_{2}$ constitutes indeed the second largest eigenvalue.

Comparing equations (\ref{eq:xl2}) and (\ref{eq:xb0}), this is the case for $N \ge 2(k+1)$, \ie for $k < N/2$. Numerical investigations suggest that the eigenvalue $\lambda_{2}$ always constitutes the second largest eigenvalue independent from the chosen values for the parameters $N$, $k$ and $q$.
However, monotonicity considerations are not that evident for $k > N/2$.

The other extremal eigenvalue $\tilde{\lambda}^{\textrm{mf}}_{\textrm{-}}$ can not be that easily assigned to a fixed index.
However, arguing similarly as for the second largest eigenvalues, it is possible to find good estimates for the index at which the smallest eigenvalue
\bel{eq:lmin}
\tilde{\lambda}^{\textrm{mf}}_{\textrm{-}} = \min_{l} \tilde{\lambda}^{\textrm{mf}}_{l} 
\ee

always occurs.

The global minimum of $f(x)$ is always located between its first two roots $x_{\text{root},1}$ and $x_{\text{root},2}$ in (\ref{eq:bpoints}), \ie

\bel{eq:xminglobal}
x_{\text{root},1}= \frac{\pi}{k + 1} < x_{\textrm{-}} <\frac{2 \pi}{k + 1}=x_{\text{root},2} \ .
\ee

It thus follows for the index $l_{\textrm{-}}$ of the smallest eigenvalue:
\be
 \frac{N}{k + 1} < l_{\textrm{-}} -1 <\frac{2 N}{k + 1} \ . 
\ee

Therefore, a good estimate for the smallest eigenvalue is given by

\bel{eq:lminestimate}
\tilde{\lambda}^{\textrm{mf}}_{\textrm{-}} \approx \tilde{\lambda}^{\textrm{mf}}_{\lfloor \frac{3 N}{2 (k + 1)}+1 \rceil} \ ,
\ee
where $\lfloor x \rceil$ denotes the nearest integer to $x$.

\section{Extreme eigenvalues}\label{extreme}

\begin{figure}[t] 
\begin{center}
\includegraphics[width=85mm]{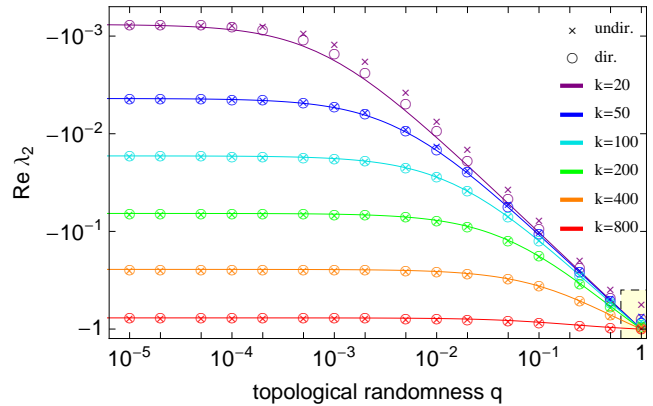}
\caption[Second largest eigenvalues from regular to random networks.]{\textbf{Second largest eigenvalues from regular to randomized networks.} Numerical measurements for undirected ($\times$) and directed ($\ocircle$) networks in comparison with the analytical mean field predictions (Eq.~(\ref{eq:mft}), solid lines) as a function of $q$, for different degrees $k$. The error bars on the numerical measurements are smaller than the data points ($N=1000$, each data point averaged over $100$ realizations). Adapted from \cite{Grabow:2012dg}.
\label{fig:lambda2vsq}
}
\end{center}
\end{figure}

As the offset of each eigenvalue (\ref{eq:lambdafinal}) equals $k$, we consider the scaled eigenvalues 
\bel{eq:scaledlambdas}
\lambda^{\textrm{mf}}_{l}(N,k,q)=\frac{\tilde{\lambda}^{\textrm{mf}}_{l}(N,k,q)}{k}
\ee 
in the following to allow for a consistent analysis for different $k$. Additionally, we always plot the real part of the eigenvalues in the case of directed networks whereas we plot $\lambda_{l} \in\mathbb{R} $ in the case of undirected networks.

As stated in the introduction the principal eigenvalue $\lambda_{*}$ (that equals $\lambda_{2}$ as shown above) is of special importance since it dominates the long time dynamics (see e.g. \cite{Grabow11}).

For $l=2$, eq. (\ref{eq:lambdafinal}) simplifies to
\begin{align}
\lambda^{\textrm{mf}}_{2}(N,k,q)=&-1 + c^\prime \left(\frac{\sin\left(\frac{(k+1)\pi}{N}\right)}{\sin\left(\frac{\pi}{N}\right)} -1\right)\nonumber \\
&+ c^{\prime\prime} \left(\frac{\sin\left(\frac{(k+1) \pi}{N}\right)}{\sin\left(\frac{\pi}{N}\right)}\right). \label{eq:mft}
\end{align}

In Fig.~\ref{fig:lambda2vsq} we compare the typical eigenvalues obtained by numerical diagonalization with our analytic prediction (\ref{eq:mft}). The analytic prediction is accurate for both undirected and directed networks, and for all but very small relative degrees $k/N<0.05$. Moreover, the prediction (\ref{eq:mft}) approximates well the actual dependence of $\lambda_2$ for all but very large $q$, thus including regular rings, small-world and even more substantially randomized network topologies. 

Next we further investigate the scaling behavior of eq. (\ref{eq:mft}) in dependence on the system's parameters.

\subsection{Approximation for small degrees}

\begin{figure}[t] 
\begin{centering}
\includegraphics[width=85mm]{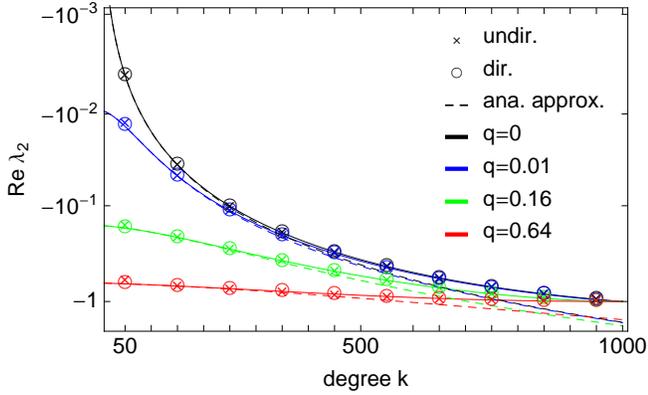}
\par\end{centering}
\caption[Fixed network size.]{{\textbf{Scaling of the real part of the second largest eigenvalue for fixed network size.} Numerical measurements for undirected ($\times$) and directed ($\ocircle$) networks in comparison with the analytical mean field predictions (Eq.~(\ref{eq:mft}), solid lines). The analytical approximations obtained from the expansions in eq. (\ref{eq:mftexpansion}) are depicted by the dashed lines. The error bars on the numerical measurements are smaller than the data points, $N$ fixed to $1000$. \label{fig:lambda2nfixedvsk}}}
\end{figure}

Expanding eq. (\ref{eq:mft}) up to $O(N^{-2})$ as $N \to \infty$ yields
\begin{align} \label{eq:mftexpansion}
\lambda^{\textrm{mf}}_{2}(N,k,q) \simeq & - q - \frac{(1 + k (1 - q)) q}{N}  \\ \nonumber
& - \frac{(k+1) (k+2) \ \pi^2 (1 - q) + 6 q (k +1 - k q)^2}{6 N^2} \ .
\end{align}

For $q=0$ we recover the known approximation for symmetric regular ring networks
\be
\lambda^{\textrm{mf}}_{2}(N,k,0) \simeq - \frac{(k+1) (k+2) \ \pi^2}{6 N^2} \label{eq:mftexpansionq0}\ .
\ee 

The approximation (\ref{eq:mftexpansion}) agrees well with eq. (\ref{eq:mft}) up to values of $k < N/2$, but still is a good guide for even larger degrees $k$, cf. Fig.~\ref{fig:lambda2nfixedvsk}.

\subsection{Scaling with network size}

In order to study the dependence on the network size we fix the edge density $d=k/N>0$ for large $N \gg 1$ which ensures that the network will remain connected. 
This leads to the expression
\begin{equation}
\lambda^{\textrm{mf}}_{2}(d,q) \simeq  -1 + \frac{(1-d)(1-q)}{(1-d (1-q))d \pi}\sin(d \pi) \label{eq:cfixed} \ 
\end{equation}
in the limit $N \rightarrow \infty$.

\begin{figure}[h!] 
\begin{center}
\includegraphics[width=85mm]{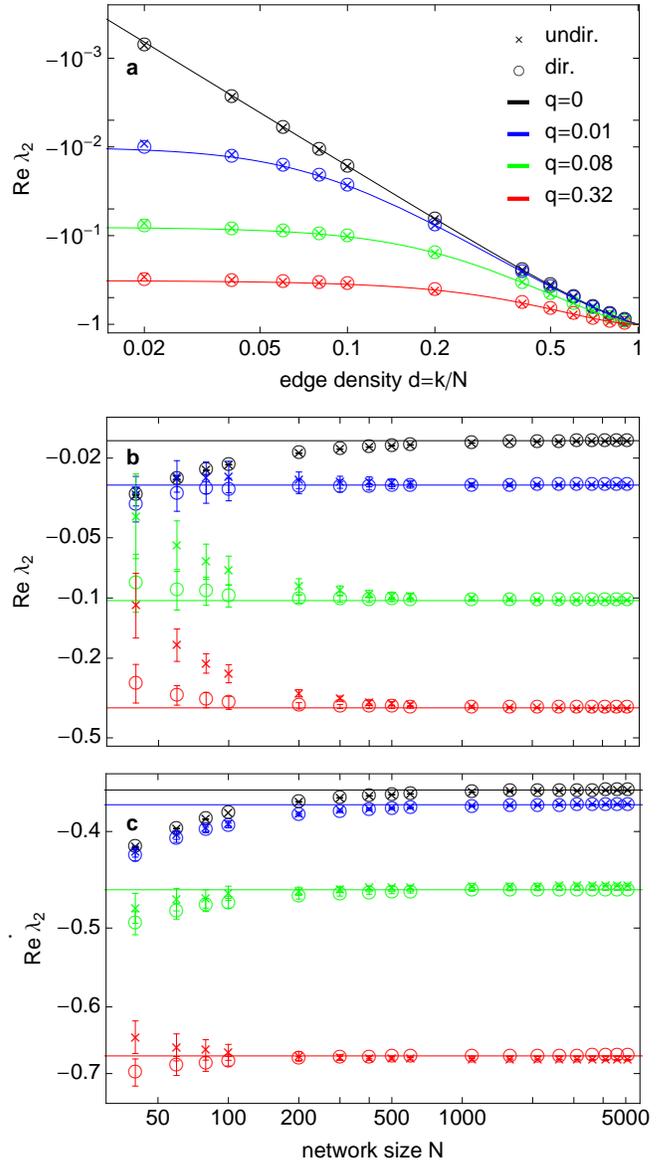}
\caption[Dependence on edge density and network size.]{\textbf{Second largest eigenvalues in dependence on edge density and network size.} \textbf{a}: Numerical measurements for directed ($\ocircle$) and undirected ($\times$) networks (error bars smaller than the data points) in comparison with the analytic mean field prediction (\ref{eq:cfixed}) for $N=2000$. 
\textbf{b}: Asymptotic ($N\rightarrow \infty$) real parts of the
second largest eigenvalues $\lambda_{2}$ in dependence on the network size $N$ for fixed edge density $d=k/N=0.1$ ($q$-values and symbols as in (\textbf{a})). 
\textbf{c}: Asymptotic ($N\rightarrow \infty$) real parts of the
second largest eigenvalues $\lambda_{2}$ in dependence on the network size $N$ for fixed edge density $d=k/N=0.5$ ($q$-values and symbols as in (\textbf{a})). \textbf{a} and \textbf{b} adapted from \cite{Grabow:2012dg}.
\label{fig:lambda2cfixed}
}
\end{center}
\end{figure}

Fig.~\ref{fig:lambda2cfixed} confirms the validity of our approximation (\ref{eq:cfixed}): the second largest eigenvalues $\lambda_{2}$, for both undirected and directed networks, in dependence on the edge density $d$ for networks of size above about $N=500$ nodes are predicted well again. Other edge densities than those displayed ($d=k/N=0.1$ and $d=0.5$, Fig.~\ref{fig:lambda2cfixed} (\textbf{b}),(\textbf{c})) qualitatively yield the same asymptotic behavior.

\subsection{The smallest eigenvalue}

The smallest eigenvalue $\lambda_{\textrm{-}}$ defined in Eq.~(\ref{eq:lmin}) is also an important indicator for synchronization properties, in particular -- in combination with the second largest eigenvalue -- for the synchronizability (see e.g. \cite{Barahona:2002bm}). For directed networks this refers to the eigenvalue with the smallest real part.

\begin{figure}[t] 
\begin{center}
\includegraphics[width=85mm]{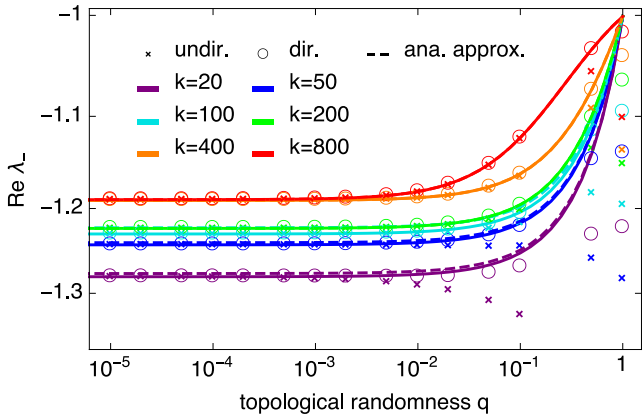}
\caption[Smallest eigenvalues from regular to random networks.]{\textbf{Smallest eigenvalues from regular to randomized networks.} Numerical measurements for undirected  ($\times$) and directed ($\ocircle$) networks in comparison with the analytical mean field predictions (Eq.~(\ref{eq:lmin}), solid lines) as a function of $q$, for different degrees $k$. Dashed lines show the analytic estimations of the smallest eigenvalues (Eq.~(\ref{eq:lminestimate})). The error bars on the numerical measurements are smaller than the data points ($N=1000$, each data point averaged over $100$ realizations).
\label{fig:lambdaminvsq}
}
\end{center}
\end{figure}

Here, the analytic prediction (\ref{eq:lambdafinal}) again fits well with the actual eigenvalues obtained by numerical diagonalization, cf.~Fig.~\ref{fig:lambdaminvsq}. Note also that our estimation (\ref{eq:lminestimate}) for the smallest eigenvalue agrees well with the actual analytic prediction (\ref{eq:lmin}). It turns out that the analytic prediction is accurate for both undirected and directed networks. The prediction (\ref{eq:mft}) approximates well the actual dependence of $\lambda_{\textrm{-}}$ for small $q$, thus including regular rings and small worlds. 
The prediction is still a good guide for the general dependence of the second largest eigenvalue on $q$, but shows some deviation from the numerical results for larger $q$, \ie for substantially randomized network topologies.

\section[Analytical predictions for random topologies]{Analytical predictions via random matrix theory}\label{RMT}

To analytically predict the second largest eigenvalues for the graph Laplacians of undirected and directed networks close to $q=1$ (see the shaded area (bottom, right) in Fig.~\ref{fig:lambda2vsq}) we consult Random Matrix Theory \cite{wigner51} (cf. also \cite{Porter65,mehta,Tao:2009wp,rmt05}).
For a review of synchronization in networks with random interactions see e.g. \cite{Feng:2006vc}.

Firstly, we consider undirected networks associated with symmetric matrices. 
Here, every connection between a pair of nodes $i$ and $j \neq i$ is present with a given probability $P$.

Secondly, we consider directed networks associated with asymmetric matrices. 
Here, all nodes have the same in-degree $k^{in}_{i}=k^{in}$. Each of the $k^{in}$ nodes that is connected to node $i$ is independently drawn from the set of all other nodes in the network with uniform probability.

Given a sufficiently large network size $N$ and a sufficiently large $k$ (respectively, a sufficiently large $k^{in}$) we numerically find that the set of non-trivial eigenvalues resemble disks of radii $r^{\prime}$ for undirected networks and $r$ for directed networks (cf. also \cite{Timme04,Timme06}). For directed networks where the in-degree $k^{in}_{i}=k$ for all nodes $i$ stays fixed during the whole rewiring procedure, \ie all diagonal elements are constant, the graph Laplacian is obtained by shifting all eigenvalues of the adjacency matrix by $-k$. For undirected networks there are small deviations from node to node but the average degree equals $k$. However, numerical simulations confirm that shifting here again the eigenvalues of the symmetric adjacency matrix by the negative average degree $-k$ is feasible.  
Thus, we consider the adjacency matrices in the following, $A^{{\textrm{sym}}}$ for undirected and $A^{\textrm{asym}}$ for directed networks and later shift them by $-k$.

\subsection{Ensembles of symmetric and asymmetric random matrices}

Firstly, consider $N\times N$ symmetric matrices $A=A^{\mathsf{T}}$ with real elements $A_{ij}$. We constrain the diagonal entries to vanish $A_{ii}=0$ and denote its $N$ eigenvalues by $\lambda_{k}$. The elements $A_{ij}$ ($i < j$) are independent, identically distributed random variables according to a probability distribution $\rho(A_{ij})$. 
According to \cite{Mirlin91,Fyodorov91,Semerjian:539372} there is only one known ensemble with independent identically distributed matrix elements that differs from the Gaussian one. Thus there are exactly two universality classes, i.e. classes which do not depend on the probability distribution $\rho(A_{ij})$, but are determined by matrix symmetry only. 
Every ensemble of matrices within one of these universality classes exhibits the same
distribution of eigenvalues in the limit of large matrices, $N\rightarrow\infty$,
but the eigenvalue distributions are in general different for the
two classes. 

The arithmetic mean of the eigenvalues is
zero,
\begin{equation}
\left[\lambda_{i}\right]_{i}:=\frac{1}{N}\sum_{i=1}^{N}\lambda_{i}=\frac{1}{N}\sum_{i=1}^{N}A_{ii}=0\label{eq:meanlambda0}
\end{equation}
and the ensemble variance of the matrix elements scales like 
\begin{equation}
\sigma^{2}=\left\langle A_{ij}^{2}\right\rangle = \frac{r^{2}}{N}\label{eq:Gaussian_variance}
\end{equation}
for $N\gg1$ and $r > 0$ being the radii of disks that enclose the set of non-trivial eigenvalues for directed networks \cite{Timme04,Timme06}.

For the Gaussian symmetric ensemble, it is known \cite{wigner51,mehta}
that the distribution of eigenvalues $\rho_{\textrm{Gauss}}^{\textrm{sym}}(\lambda)$
in the limit $N\rightarrow\infty$ is given by Wigner's semicircle
law 
\bel{eq:wigner}
\rho_{\textrm{Gauss}}^{\textrm{sym}}(\lambda)=\left\{ \begin{array}{ll}
\frac{1}{2\pi r^{2}}\sqrt{4r^{2}-\lambda^{2}}  & \textrm{if }|\lambda|\leq2r\\
0 & \textrm{otherwise}.\end{array}\right.
\ee
The ensemble of sparse matrices \cite{Bray88,Rodgers:1988vv,Fyodorov91,Mirlin91,Rodgers:2005dn,Goetze2010}
exhibits a different eigenvalue distribution $\rho_{\textrm{sparse}}^{\textrm{sym}}(\lambda)$
that depends on the finite number $k$ of nonzero entries per row
and approaches the distribution $\rho_{\textrm{Gauss}}^{\textrm{sym}}(\lambda)$
in the limit of large $k$ such that\begin{equation}
\lim_{k\rightarrow\infty}\rho_{\textrm{sparse}}^{\textrm{sym}}(\lambda)=\rho_{\textrm{Gauss}}^{\textrm{sym}}(\lambda).\label{eq:sparseGauss}\end{equation}
It is important to note that in the limit of large $N$ the eigenvalue distributions
$\rho_{\textrm{sparse}}^{\textrm{sym}}$ and $\rho_{\textrm{Gauss}}^{\textrm{sym}}$ depend only on the one parameter $r$, that is derived
from the variance of the matrix elements (\ref{eq:Gaussian_variance}).

For real, asymmetric matrices (independent $A_{ij}$ and
$A_{ji}$), there are no analytical results for the case of sparse
matrices but only for the case of Gaussian random matrices. The Gaussian
asymmetric ensemble yields the distribution of complex eigenvalues
in a disk in the complex plane \cite{Girko85,Sommers:1988uq}
\begin{equation}
\rho_{\textrm{Gauss}}^{\textrm{asym}}(\lambda)=\left\{ \begin{array}{ll}
\frac{1}{\pi r^{2}} & \textrm{if }|\lambda|\leq r\\
0 & \textrm{otherwise}\end{array}\right.\label{eq:uniform_EW_distribution}
\end{equation}
where $r$ from Eq.\ (\ref{eq:Gaussian_variance}) is the radius
of the disk that is centered around the origin. Like in the case of symmetric
matrices, this distribution also depends only on one parameter $r$,
that is derived from the variance of the matrix elements. 

\subsection{Undirected random networks}

The real symmetric adjacency matrix $A^{{\textrm{sym}}}$ is an $N\times N$ matrix that satisfies $A^{\textrm{sym}}_{ij}=A^{\textrm{sym}}_{ji}$ and $A^{\textrm{sym}}_{ii}=0$.

Furthermore, the matrix elements of $A^{{\textrm{sym}}}$ are independent up to the symmetry constraint $A^{\textrm{sym}}_{ij}=A^{\textrm{sym}}_{ji}$. They are equal to $1$ with
probability 
\be
P=\frac{\left\langle k_{i} \right\rangle}{N-1}\approx \frac{k}{N} \ ,
\ee
and equal to $0$ with probability $1-P$.

Thus, the variance $\sigma^{2}$ is given by
\bel{eq:sigma}
\sigma^{2}=P(1-P)=\frac{k}{N}(1-\frac{k}{N}) \ .
\ee

Therefore, the eigenvalues are located in a disc of radius 
\be r^{\prime}=2 r
\ee 
with
\begin{equation}
r =  \sigma \sqrt{N} =\sqrt{k-\frac{k^{2}}{N}}  \label{eq:rundir}
\end{equation}
centered around the origin.

\subsection{Directed random networks}

The real asymmetric adjacency matrix $A^{\textrm{asym}}$ has exactly $k$ elements equal to one per row. Therefore, its elements have a spatial average
\begin{equation}
\big[A^{\textrm{asym}}_{ij}\big]:=\frac{1}{N}\sum_{j=1}^{N}A^{\textrm{asym}}_{ij}=\frac{k}{N}
\end{equation}
and a second moment
\be
\left[({A_{ij}^{\textrm{asym}}})^{2}\right]=\frac{1}{N}\sum_{j=1}^{N}{(A_{ij}^{\textrm{asym}})}^{2}=\frac{k}{N} \ .
\ee
Thus, the variance 
\be
\sigma^{2}=\left[{(A_{ij}^{\textrm{asym}})}^{2}\right]-\left[A^{\textrm{asym}}_{ij}\right]^{2} =\frac{k}{N}-\frac{k^{2}}{N^{2}} \ .
\ee 

If we assume that the eigenvalue distribution for directed networks with fixed in-degree is similar to those for random matrices \cite{Timme04,Timme06}, we obtain a prediction 
from Eq.~(\ref{eq:Gaussian_variance}), which yields
\begin{equation}
r = \sigma \sqrt{N}= \sqrt{k-\frac{k^{2}}{N}} \label{eq:rdir}
\end{equation}
for the radius of the disk of eigenvalues centered around the origin.

\subsection{Predictions for the scaled graph Laplacians}

\begin{figure}[t]
\begin{center}
\includegraphics[width=85mm]{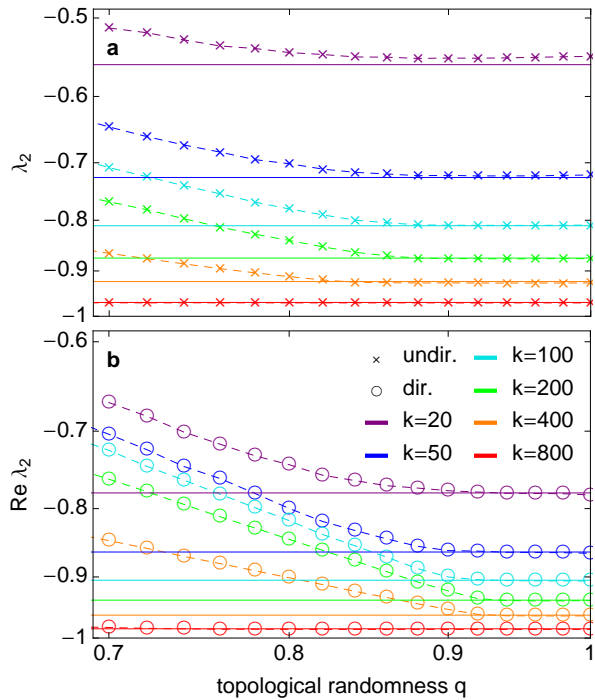}
\caption[Second largest eigenvalues for random networks.]{\textbf{Analytic prediction of the second largest eigenvalues close to $q=1$.} \textbf{a}: Numerical measurements for undirected ($\times$) networks in comparison with the analytical predictions $\lambda^{\textrm{wsc}}_{2}$ via Wigner's semi-circle law (Eq.~(\ref{eq:wsc}), solid lines), for different degrees $k$. \textbf{b}: Numerical measurements for directed ($\ocircle$) networks in comparison with the analytical predictions $\lambda^{\textrm{rmt}}_{2}$ from the theory of asymmetric random matrices (Eq.~(\ref{eq:rmt}), solid lines). The error bars on the numerical measurements are smaller than the data points ($N=1000$, each data point averaged over $100$ realizations). Dashed lines are only a guide to the eye. Taken from \cite{Grabow:2012dg}.
\label{fig:lambda2q1}
}
\end{center}
\end{figure}

To obtain predictions for the eigenvalues of the appropriate graph Laplacian, we have to consider the shift by $-k$ (discussed in the beginning of this section) and the scaling factor $1/k$ introduced in eq.~(\ref{eq:scaledlambdas}).

Together with eq.~(\ref{eq:rundir}), the second largest eigenvalues for undirected networks close to $q=1$ (Fig.~\ref{fig:lambda2q1}, (\textbf{a})) are well predicted by Wigner's semi-circle law (wsc):

\begin{align} \label{eq:wsc} 
\lambda^{\textrm{wsc}}_{2}(N,k,1)&=\frac{1}{k} \left(2 \sqrt{k-\frac{k^{2}}{N}} - k\right)  \nonumber \\
&=2 \sqrt{\frac{1}{k} - \frac{1}{N}} - 1 \ . 
\end{align}
The real parts of the eigenvalues for directed networks close to $q=1$ (Fig.~\ref{fig:lambda2q1}, (\textbf{b})) with eq.~(\ref{eq:rdir}) are with the theory of asymmetric random matrices (rmt) 
\begin{align} \label{eq:rmt}  
\lambda^{\textrm{rmt}}_{2}(N,k,1)&=\frac{1}{k} \left(\sqrt{k-\frac{k^{2}}{N}} - k\right) \nonumber \\
&=\sqrt{\frac{1}{k} - \frac{1}{N}} - 1 \ .
\end{align}

Note that $\lambda^{\textrm{wsc}}_{2}(N,k,1)$ in eq.~(\ref{eq:wsc}) acquires a positive value for too small $k$-values and a sufficiently large network size $N$ (cf.~\cite{Farkas:2001ig}). However, for the $k$-values we investigated (Fig.~\ref{fig:lambda2q1}, (\textbf{a}) and (\textbf{b})), the second largest eigenvalues are well predicted by both eqs.~(\ref{eq:wsc}) and (\ref{eq:rmt}).

\section{Summary and discussion} \label{sum2}

In this article we have presented and explicated derivations
and extended a simple mean-field rewiring scheme
suggested recently \cite{Grabow:2012dg} to derive analytical predictions for the spectra of graph Laplacians. 
The key is replacing a stochastic realization of a rewired network at given
topological randomness by its ensemble-averaged network. We achieve this
averaging via a two-stage approach that distinguishes the original outer
ring structure and the originally 'empty' inner part of a network and rewiring probabilities separately. For all $q$, the
resulting average network in particular shares exactly the same (average)
fraction of links in the original regular part of the network as well as in
its originally 'empty' part. We derive expressions for the largest
nontrivial and the smallest eigenvalues, the full spectrum, as well as several scaling behaviors. 

We remark that on theoretical grounds, the eigenvalue spectrum of the
resulting average network in general is \emph{not} equal to the average of the
spectra of the individual stochastic network realizations.
Yet, systematic numerical checks confirm that the mean field approximation
introduced is accurate as long as $q$ is sufficiently below one. In the limit
$q\rightarrow 1$, we derive the eigenvalue spectra based on random matrix
theory which become exact in the limit of infinitely large networks,
$N\rightarrow \infty$.

Although the mean field rewiring is undirected, eigenvalues for directed networks are approximated more accurately and in a wider range of $q$-values, which is in particular related to the fact that the predictions for the undirected second-largest eigenvalues at $q=1$ are larger in real part than the directed ones, while all the mean field eigenvalues converge to $-1$ at $q=1$. For `small' $k$-values the mean field approximation becomes less accurate, which may be due to the fact that the ring structure is destroyed more easily while rewiring. Additionally, the bulk spectra spread much more drastically with $q$ than for larger $k$-values.

Furthermore, note that the analysis of the mean field spectrum presented here
can principally not be extended to the Laplacian eigenvectors as these are
independent of the mean field Laplacian's elements $c_{i}$
\ref{eq:eigenmatrixelements}, just because of the circulant structure of the
mean field graph Laplacian. Consequently, the eigenvectors are the same and
always non-localized, independent of the system's parameters $N$, $k$ and $q$.
Studying distributed patterns of relaxation processes and potential
localization phenomena thus requires access to eigenvectors beyond the mean
field approximation. Studies of the Laplacian eigenvectors are rare although there are fascinating results as well. For instance, the discrete analogs of solutions of the Schr\"odinger equation on manifolds can be investigated on graphs (cf.~e.g.~\cite{Biyikoglu:2007ui}).

The simple mean field approach presented above still substantially reduces
computational efforts when studying randomized (regular or small-world)
network models. 

Generalizing our mean field approach to higher dimensions and/or to other rewiring approaches, as for
instance, relevant for neural network modeling \cite{Sporns2009}, it will serve as a powerful tool to gain new insights into the relations between structural and dynamical properties of complex networks. 

Supported by the BMBF, Grant No. 03SF0472A-E (CG,MT,JK), by a grant of the Max Planck Society (MT) and by an EPSRC Grant No. EP/E501311/1 (SG).

%\bibliography{SWspectrachaosv3}

\end{document}